**Title:** Collaboration Diversity and Scientific Impact

**Authors:** Yuxiao Dong, Hao Ma, Jie Tang, Kuansan Wang

Email: yuxdong@microsoft.com; haoma@microsoft.com; jietang@tsinghua.edu.cn;
To whom correspondence should be addressed: *Kuansan Wang,* kuansanw@microsoft.com

**Affiliations:**
YD, HM, and KW: Microsoft Research, Microsoft Corporation, One Microsoft Way, Redmond, WA 98052, USA;
JT: Department of Computer Science and Technology, Tsinghua University, Beijing 10084, China;
HM: HM is currently at Facebook Inc and this work was done when HM was at Microsoft Corp.

**Abstract:**
The shift from individual effort to collaborative output has benefited science, with scientific work pursued collaboratively having increasingly led to more highly impactful research than that pursued individually. However, understanding of how the diversity of a collaborative team influences the production of knowledge and innovation is sorely lacking. Here, we study this question by breaking down the process of scientific collaboration of 32.9 million papers over the last five decades. We find that the probability of producing a top-cited publication increases as a function of the diversity of a team of collaborators---namely, the distinct number of institutions represented by the team. We discover striking phenomena where a smaller, yet more diverse team is more likely to generate highly innovative work than a relatively larger team within one institution. We demonstrate that the synergy of collaboration diversity is universal across different generations, research fields, and tiers of institutions and individual authors. Our findings suggest that collaboration diversity strongly and positively correlates with the production of scientific innovation, giving rise to the potential revolution of the policies used by funding agencies and authorities to fund research projects, and broadly the principles used to organize teams, organizations, and societies.

**Main Text:**
Studies show that in the past century, the narrative of science has gradually shifted from the landmark contributions delivered by individual scientists on their own---such as those by Einstein in Physics, Turing in Computer Science, and Nash in Economics---to scientific advances made by team work across different fields [1] [2] [3]. In 2007, the anatomy of digitalized scientific publications conducted by a Northwestern University team---Wuchty, Jones, and Uzzi---suggests that team collaborations are more likely to produce highly cited research than individual efforts [1]. Further research has also suggested that team size generally plays a positive effect on a publication's citation counts [1] [4]; the bigger the team, the greater the scientific impact as measured by citation counts. Moreover, the follow-up work by the same team on 4.2 million papers from 662 U.S. universities further points out that collaborations have not only happened within the same organizations, but have also expanded across geographically different institutions [2]. A very recent example can be seen from the LIGO Scientific Collaboration (LSC), a team made up of scientists from more than 100 institutions that engenders a collaborative effort between the three 2017 Nobel Laureates in Physics: Weiss from Massachusetts Institute of Technology, and Barish and Thorne from California Institute of Technology, the three of whom collectively account for "decisive contributions to the LIGO detector and the observation of gravitational waves." [5]

Notwithstanding the driving force of multi-university collaborations on the production of scientific knowledge and innovations, what influence the organizations of a collaborative team have on scientific production is much less well-understood. For example, literature does not answer the question of how the number of organizations involved in a publication affects the dissemination of its scientific impact. When the size of a publication's author list is controlled for (e.g., four authors), is a collaboration between all authors from the same institution more likely to collect more citations than the case where each of the authors is affiliated with different institutions? Furthermore, when the number of authors and affiliations are fixed for a given publication (e.g., where a publication's four authors are affiliated with two institutions), do collaborations between, say, two institutions each with two authors generally produce higher-impact research than collaborations between three authors from one institution with the remaining author from the other?

In this study, we show that the way in which a collaboration is assembled across different institutions influences its scientific output. Our analysis is performed on the publicly available Microsoft Academic Graph dataset [3] comprised of more than 173 million research publications as of April 2018. Our focus is placed on the 33 million published between 1965 and 2015, which are fully associated with authors' affiliation information. Similar to previous studies [1] [3], this big academic data suggests a growing shift toward team collaborations in scientific production, with over 74% of the publications in this century produced by collaborative efforts, approximately one-and-a-half times the ratio of those from about five decades ago in the 1960s (See Fig. 1).

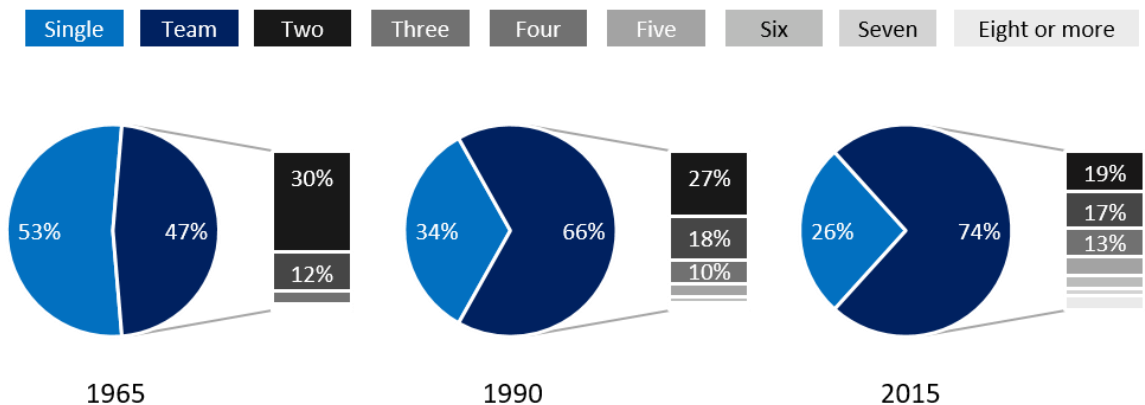

**Fig. 1. The increasing trend of bigger collaboration teams in science.** Each color represents the size of scientific collaboration teams, that is, the number of authors in a paper. The corresponding proportion represents the ratio of papers that were authored by different sizes of teams at each year.

Uniquely, we show that among the publications produced by teams, 63% of those in the 1960s were collaborations between two scientists. Since then, collaborations have consistently grown to encompass larger and larger teams, with 75% of those in this century formed between three or more people, leaving two-author publications to comprise only one quarter of published team work. Different from Wuchty et al.'s treatment [2], we further anatomize multi-institution collaborations in Fig. 2, with each circle denoting an author and each polygon representing the covered authors from the same institution. We observe universal increases of all different types of multi-institution collaborations over the past five decades when the author list size is set to between two and six, suggesting a diversification process of scientific collaborations between institutions over the past five decades. A natural question that arises here is: How do the diverse institutional collaborations affect the production of science?

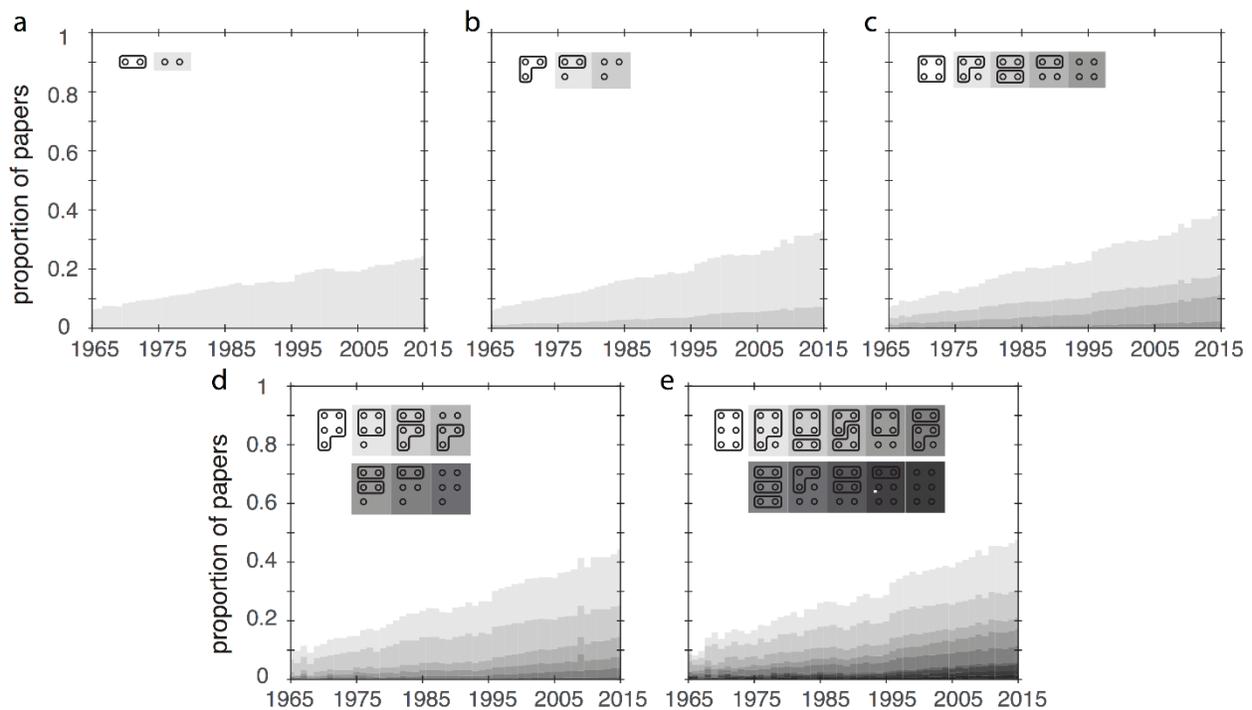

**Fig. 2. The increasing trend of multi-institution collaborations of teams with size between two and six.** Each circle denotes an author and each polygon with circles inside represents that the covered authors are from the same institution. One individual circle without a polygon covered represents that the circle's corresponding author has a distinct affiliation with his or her collaborators.

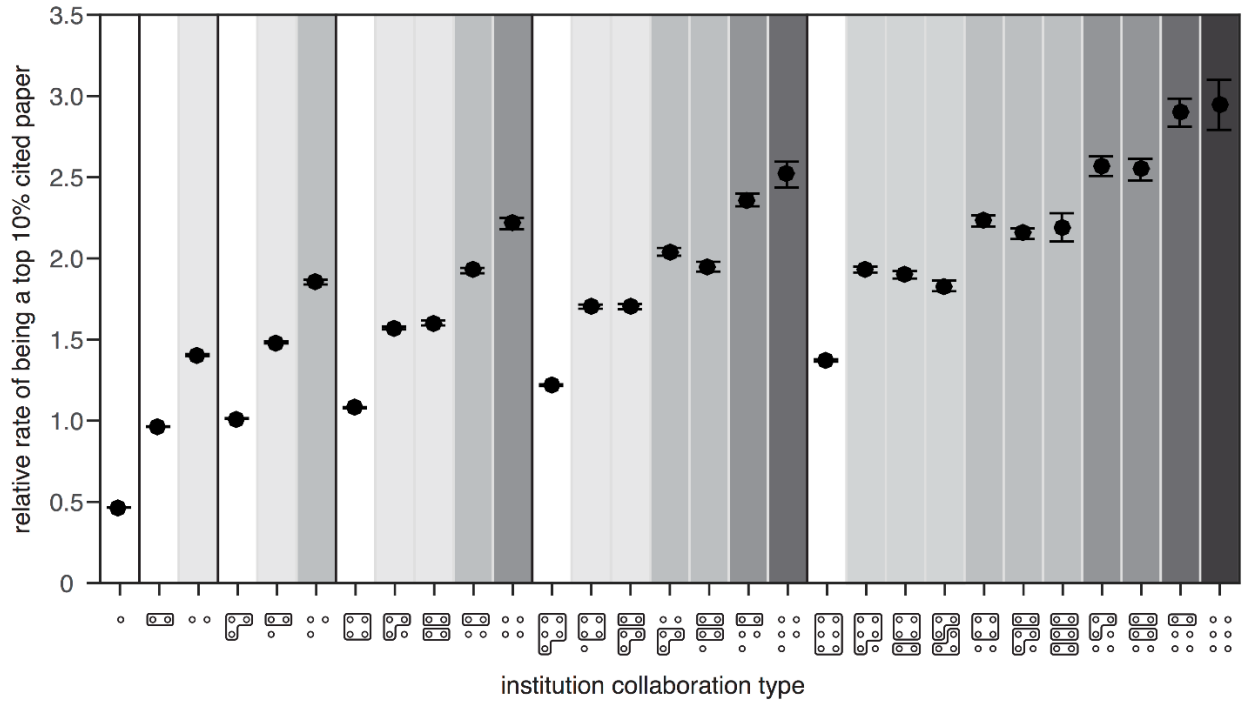

**Fig. 3. Institution collaboration diversity correlates with scientific impact.** The relative rate of being a top 10% cited paper conditioned on institutional collaboration types for one- to six-author teams. Error bars represent 95% confidence intervals, which implicitly reveal the relative frequency of the different institution collaboration types.

To answer the question, we investigate the interplay between citation counts and collaboration types of papers published between 1965 and 2015. In particular, we estimate the probability that as of 2018, a paper is among the top $k$% cited ones of those published in the same year, conditioned on the collaboration type of its authors' affiliated institutions. If institutional collaboration types do not account for the production of scientific knowledge, papers with different collaboration types would be expected to have equal chances (i.e., $k$%) to label as world-leading innovations. In Fig. 3, we report the relative rates of papers being 10% most-cited ($k=10$) for collaborations with team size between one and six (See Fig. S1 for seven-author team collaborations), where 1.0 indicates the papers are exactly 10% likely to be among the top 10% cited.

In general, it is immediately observable that institution collaboration types play a significant role in the dissemination of a publication's scientific impact. Remarkably, we observe a clear stratification in the probabilities of being a top-cited research work based on the number of institutions involved. More strikingly when the number of institutions involved in a fixed-size collaboration is controlled for, different collaboration types offer relatively similar potential to generate high-impact science, such as the three-one and two-two two-institution collaborations for four authors, and the four-one-one, three-two-one, and two-two-two three-institution collaborations for five authors. Overall, if we fix the size of a paper's author list, then as the number of institutions participating in the research increases, the relative rate of being a top-cited research publication also increases.

We refer to the number of institutions represented in a paper as a measure of collaboration diversity. When forging a collaboration across multiple institutions, authors from the same institution---who are considered likely to share similar knowledge and perspectives---are expected to contribute an individual and unique piece of creative ideas to the collaboration. Therefore, with more institutions involved into a team, more diverse creativity is inserted into a scientific project, possibly fostering more exceptional innovations in science. Indeed, our examination on this big academic dataset suggests that collaboration diversity positively correlates with the chances of generating high-impact research across different team sizes in a linear manner.

To quantitively measure the effect, we formalize the concept of the collaboration diversity impact index $I_m^k$ for teams with size $m$. $I_m^k$ is defined as the ratio between the rate of being a top $k\%$ cited paper produced by the most diverse collaborations---wherein each author comes from one distinct institution---and the rate of being one produced by the least diverse team---where collaborations happen within the same institution. An $I_m^k$ equal to 1 indicates that collaboration diversity does not play a role in the dissemination of scientific impact. When $I_m^k$ is larger than 1, it indicates that collaboration diversity is positively correlated with the probability of producing high-impact research, and vice versa when $I_m^k$ is lower than 1. Observed from Fig. 3, the collaboration diversity impact indices $I_{2\sim6}^{10}$ for teams with size from two to six are 1.46, 1.83, 2.05, 2.06, and 2.15, respectively, representing an increasingly positive effect of collaboration diversity as teams become larger.

Conventional wisdom suggests that the processes of producing scientific knowledge may vary across disciplines. To this end, we take a similar treatment with Wuchty, Jones, and Uzzi's studies [1] [2], and group the 33 million publications into three sets based on their fields of studies: Science (covering Physics, Chemistry, Biology, etc.), Applied Science & Engineering (covering Computer Science, Electronic Engineering, etc.), and Social Sciences (covering Sociology, Philosophy, Sociology, etc.). Remarkably, we observe similarly positive correlations between institution diversity and scientific impact in different groups (See Fig. S2—S4). For example, the collaboration diversity impact indices $I_4^{10}$ are 1.72, 2.24, and 1.72 in Sciences, Applied Sciences & Engineering, and Social Sciences, respectively, suggesting the universally positive correlation between collaboration diversity and the generation of high-impact research.

When applied to scientific work, the Matthew effect posits that the inclusion of a highly reputable institution into a team may increase the visibility of the work [6, 2], potentially leading to higher scientific impact as measured by the number of citations attracted. To this end, we further anatomize multi-institution collaborations from the perspective of institutional stratification. In specific, we follow the same treatment in [2] to group all academic institutions into four tiers based on their rankings, with those ranked in the top 5% as tier I, top 6—10% as tier II, top 11—20% as tier III, and the remaining as tier IV. Their rankings are based on the total citation counts collected by their single-institution publications published between the years 1965 and 2015. Fig.

4a presents the relative rates that different institution-tiers of three-author collaborations produce top 10% impactful scientific work (See Fig. S5—S9 for different sizes of teams). It is clear that the positive correlation between collaboration diversity and scientific impact can be universally observed within the same tier and across different tiers of institutional collaboration. Strikingly, we also notice that diverse institutional collaborations can enable lower-tier institutions to generate more impactful science than higher-tier ones. For example, it is more likely for three authors from three tier-II institutions to produce top-cited research than three from one single tier-I institution.

To minimize the effect of authors' scientific impact on the discovery, we further fix the tier of institutions involved in collaborations and stratify the authors. For example, we rank all authors in tier-I institutions based on their citation count collected until the year 2009 and group them into four tiers using the same splitting thresholds above. We then examine how collaborations between different tiers of authors affect the influence of institutional collaboration diversity by using papers published between the years 2010 and 2015. Fig. 4b demonstrates that when both the tiers of institutions and authors are controlled for, collaboration diversity still demonstrates a positive interplay with the dissemination of scientific impact in each tier of three-author collaborations. For other size of teams, see Fig. S10—S14.

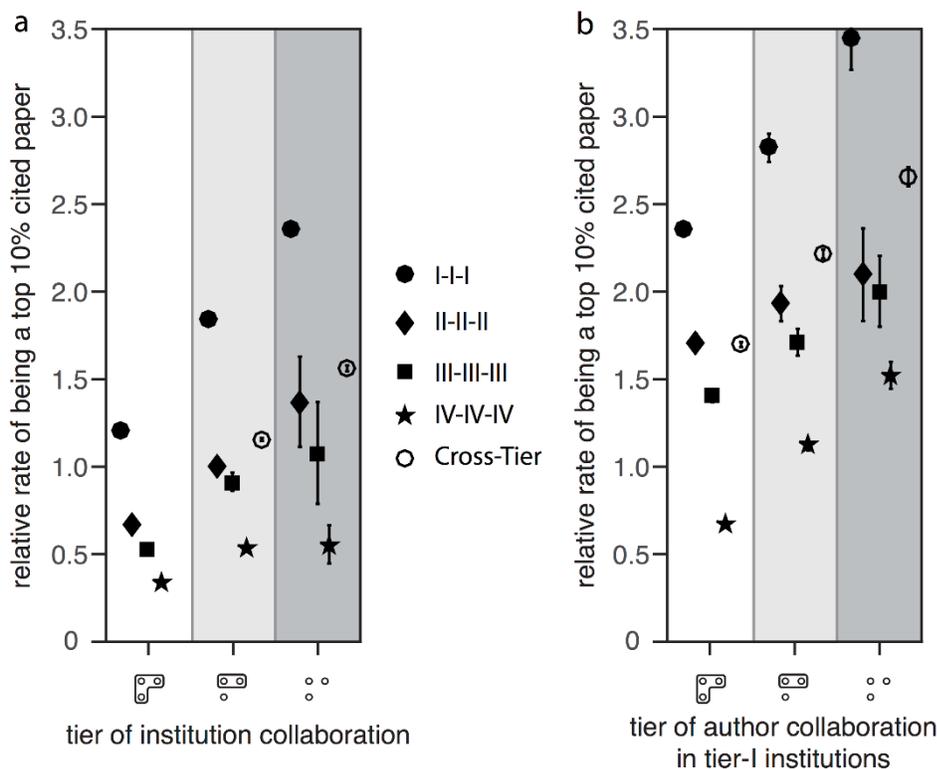

**Fig. 4. Tiers of both institution and author collaboration correlate with scientific impact.** X-X-X indicates that the three collaborating institutions (a) or authors (b) are from the same tier X, where X represents I, II, III, or IV. Cross-tier represents that the collaboration teams are formed with different tiers of institutions (a) or authors (b). In specific, all authors in (b) are from the tier-I institutions.

One may suspect that papers representing collaborations between multiple institutions could benefit from the self-citation behavior [1]. To assess its influence on our discoveries, we remove all self-citation relationships between papers. The resultant collaboration diversity indices $I_{2\sim6}^{10}$ for two- to six-member teams are 1.42, 1.79, 2.02, 2.01, and 2.09, respectively, suggesting the relatively consistent effects of collaboration diversity on the production of high-impact research in both the presence and absence of self-citation behavior (See Fig. S15).

The other commonly asked question is how different time-frames may influence the dissemination of scientific impact. Therefore, in addition to using citations collected to date (i.e., as of 2018), we also look at only those citations that were accrued within different durations since the publication year of the papers. For all papers published between 1965 and 2008, the collaboration diversity impact indices $I_{2\sim6}^{10}$ counting only citations collected within five years of publication are 1.30, 1.57, 1.84, 1.83, and 1.92, all of which are clearly larger than 1, but not as strong as when counting all citations collected to date, i.e., 1.50, 1.83, 2.08, 2.08, and 2.12 (See Fig. S16 and Fig. S18). Within a longer time-frame, e.g., ten years, the corresponding indices $I_{2\sim6}^{10}$ are 1.38, 1.68, 1.92, 1.92, and 1.95, which are very close to the results when counting all citations to date (See Fig. S17—S18). As longer time-frames are given for disseminating scientific impact, collaboration diversity impact indices gradually increase, suggesting the amplified effect of collaboration diversity through the test of time.

So far, we have studied the role of collaboration diversity in the citations collected for papers published between the year 1965 and 2015 (2008) as a whole. However, as the volume of publications has grown at an exponential rate over the last century [3] [7], the observed effect of collaboration diversity may be dominated by recent publications. Hereby we split the studied publications into several groups based on their publication dates. Taking the collaboration diversity impact index $I_4^{10}$ as an example, the data shows that it was 1.56 in the 1960s and 1970s, 1.83 and 1.81 in the following decades, 2.25 in the first decade of this century, and 2.04 between 2010 and 2015, suggesting the universal patterns between collaboration diversity and scientific impact across generations.

By far, our study has focused on the production of exceptional innovations as measured by being a top *k*% (k=10) most-cited paper among those published at the same year. We further examine whether the discoveries vary according to the change of the threshold *k*. We find that increasing the bar of outstanding science, i.e., narrowing down the percentage *k*, remarkably increases the collaboration diversity impact indices (See Fig. S19—S24). For example, the indices $I_{2\sim6}^{1}$ for top 1% cited-work reaches 1.96, 2.92, 3.79, 4.30, and 5.79, respectively, which are significantly larger than the corresponding cases for top 10% that ranges between 1.46 and 2.15. Generally, as the bar decreases---that is, as *k* increases---the correlation between collaboration diversity and scientific impact declines. Notice that though $I^{100}$ (k=100) is in theory guaranteed to converge to 1, $I^k$ could be either larger or smaller than 1 when *k<100*. That said, our studies reveal that collaboration diversity correlates most favorably with the production of world-leading scientific innovations (e.g., top 1% cited research).

Recall that studies have observed that team size has a positive effect on papers' future citation counts. Surprisingly, however, we find that if collaboration diversity is accounted for, smaller but more diverse teams are more likely to produce world-leading innovations than relatively larger but less-diverse collaborations. In fact, it is more likely for collaborations between two authors from two distinct institutions to generate highly impactful research than larger teams with three-to-five members who are from a single organization (See Fig. 1). Similarly, for collaborations between three and six authors, the most diverse teams---with all members from different organizations---are 50% more likely to produce top 10% most-cited research than the most homogeneous teams with one additional member. The phenomenon here suggests that the positive effect of team size on research impact can often be an oversimplification, and the diversity of team members should also be considered when forming a collaboration.

The digitally recorded scientific publications reveal the increasing dominance of diverse teams in the production of world-leading innovations over the past half century. The observation provides us with a better understanding of institutional collaborations in scientific production, as well as strong empirical evidence in support of collaborative research. For example, our results align with the widely accepted academic norms that universities tend to not hire their own students as faculty after their immediate graduation, forging the inclusive environment to incorporate fresh ideas and diverse experiences from others into their own systems. In fact, in 2016, the National Science Foundation (NSF) of the United States released the NSF INCLUDES initiative to call for diversity and collaborative proposals in research project funding [8]. Moreover, the result can be also generalized to academic enrollment as well as workplace employment. Broadly, the discoveries offer support for fundamental principles by which to organize teams, organizations, and societies.

**Acknowledgements**: The authors would like to thank Reid A. Johnson and Brian Uzzi for their insightful comments and suggestions.

**Supplementary Information** for Collaboration Diversity and Scientific Impact

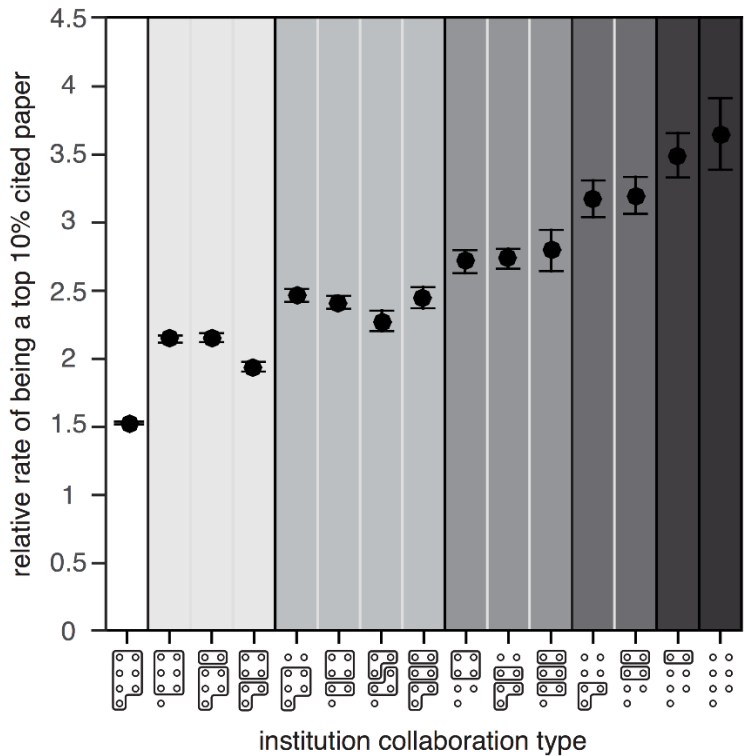

**Fig. S1. Seven-author institution collaboration diversity correlates with scientific impact.** The relative rate of being a top 10% cited paper conditioned on institutional collaboration types for seven-author teams.

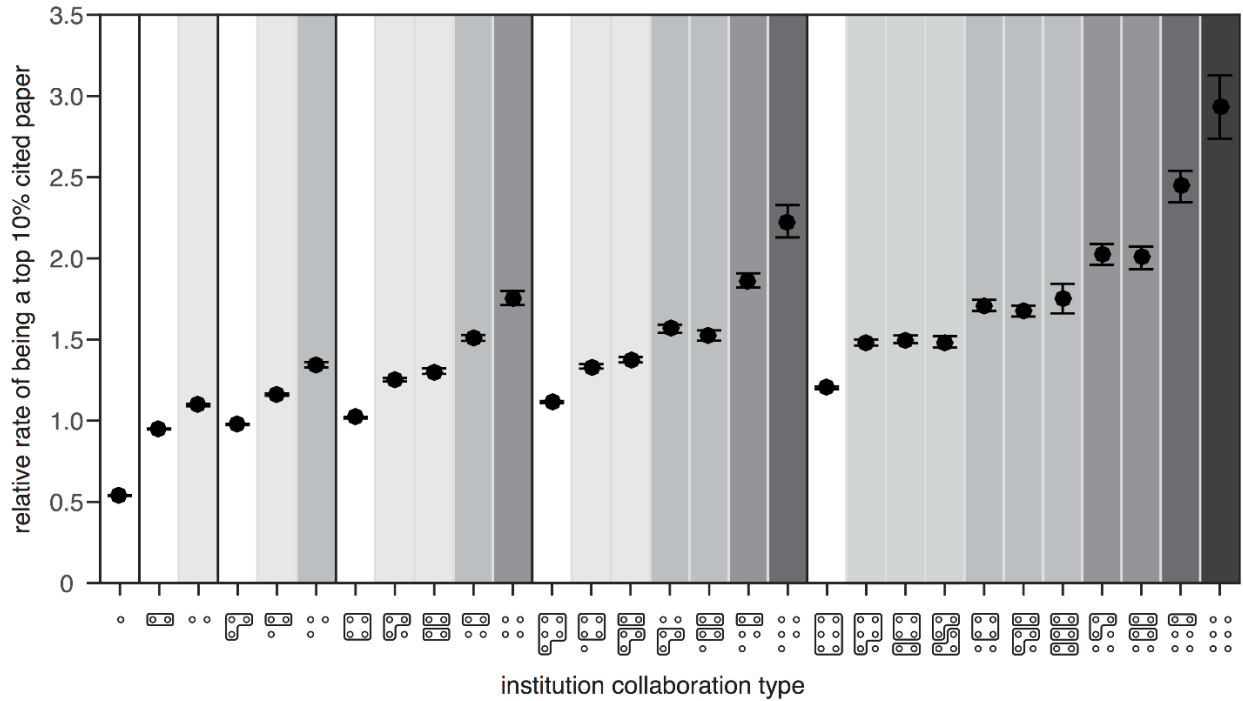

**Fig. S2. Institution collaboration diversity correlates with scientific impact in the fields of science.** The relative rate of being a top 10% cited paper conditioned on institutional collaboration types for one- to six-author teams.

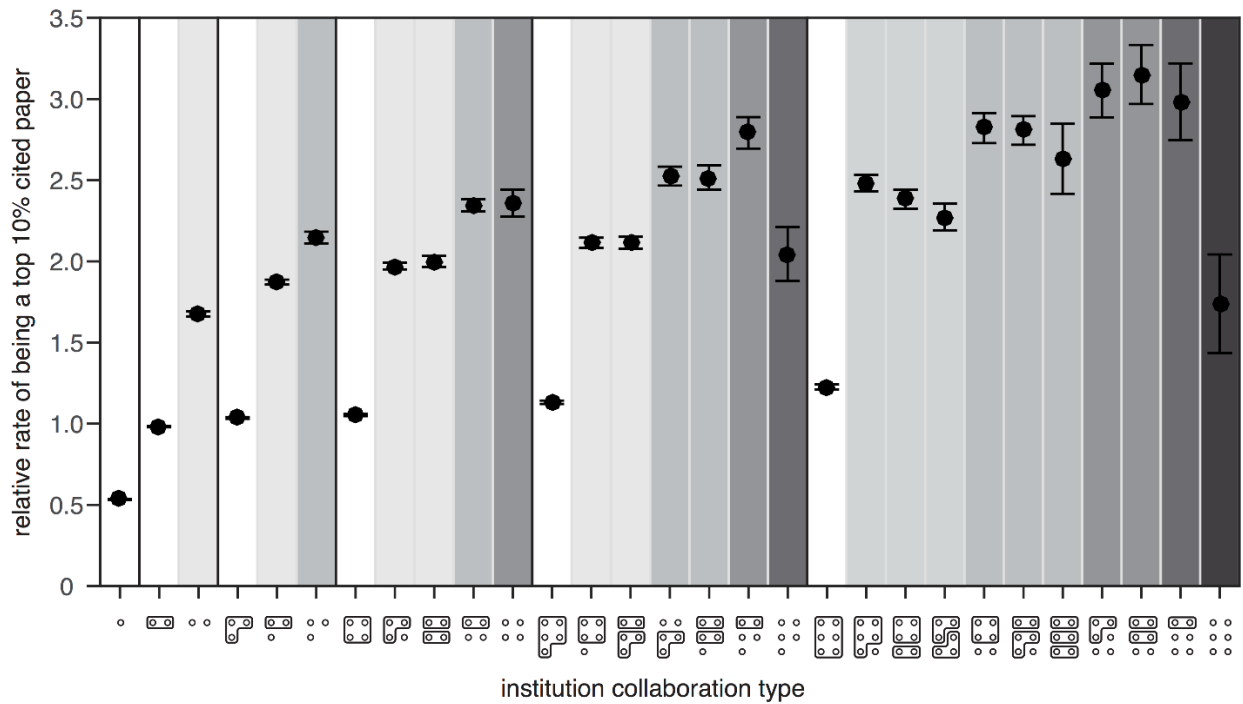

**Fig. S3. Institution collaboration diversity correlates with scientific impact in the fields of applied sciences and engineering.** The relative rate of being a top 10% cited paper conditioned on institutional collaboration types for one- to six-author teams.

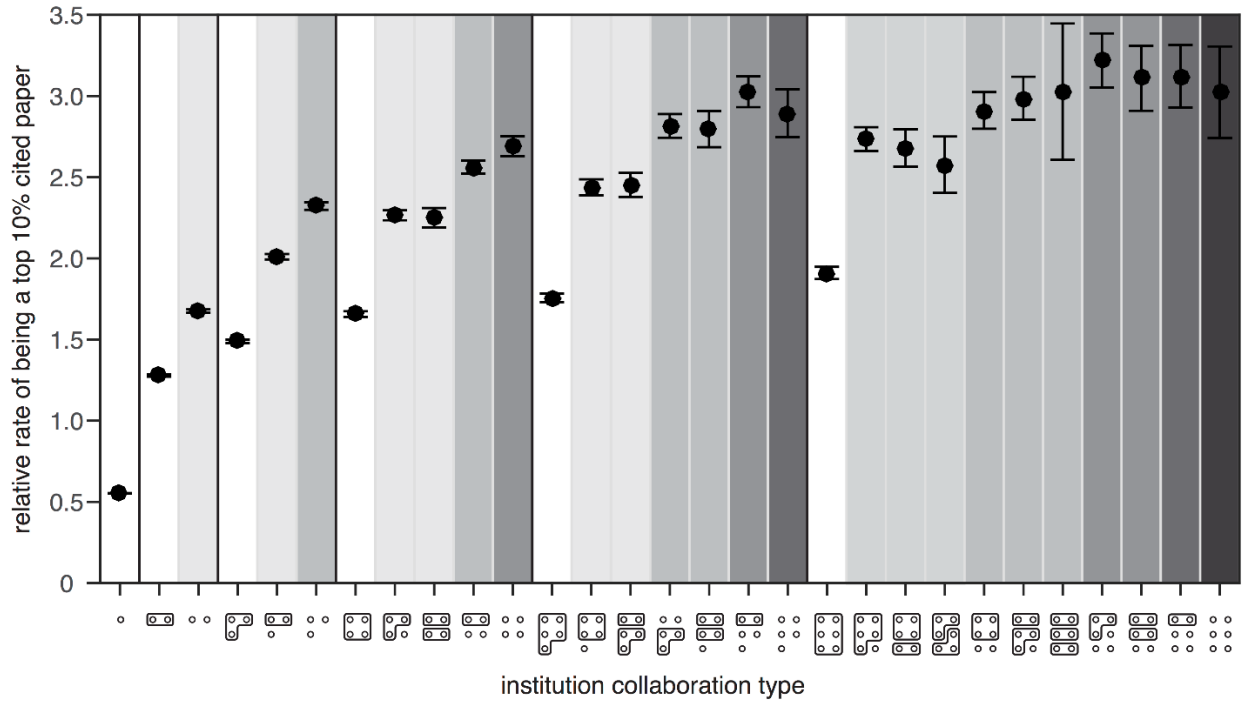

**Fig. S4. Institution collaboration diversity correlates with scientific impact in the fields of social sciences.** The relative rate of being a top 10% cited paper conditioned on institutional collaboration types for one- to six-author teams.

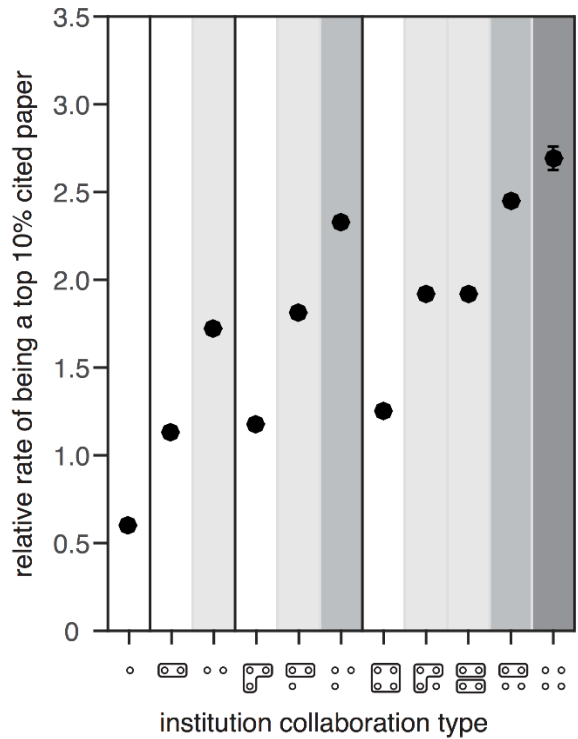

**Fig. S5. Tier-I institution collaboration correlates with scientific impact.**

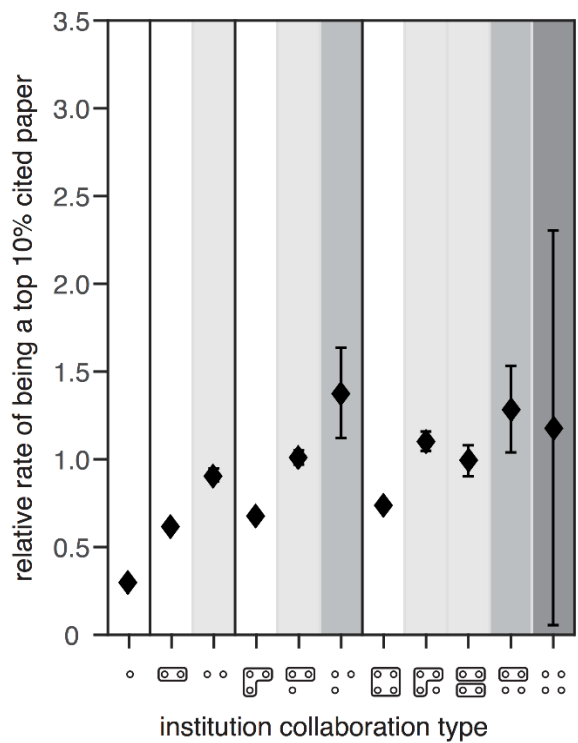

**Fig. S6. Tier-II institution collaboration correlates with scientific impact.**

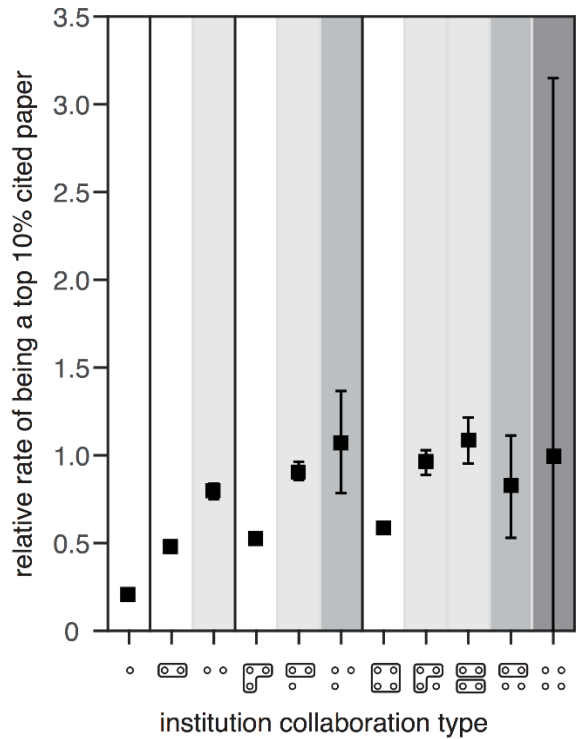

**Fig. S7. Tier-III institution collaboration correlates with scientific impact.**

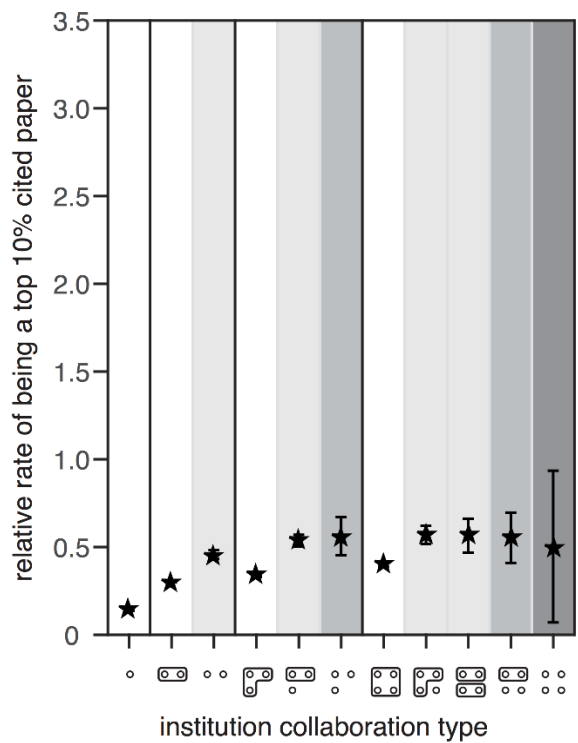

**Fig. S8. Tier-IV institution collaboration correlates with scientific impact.**

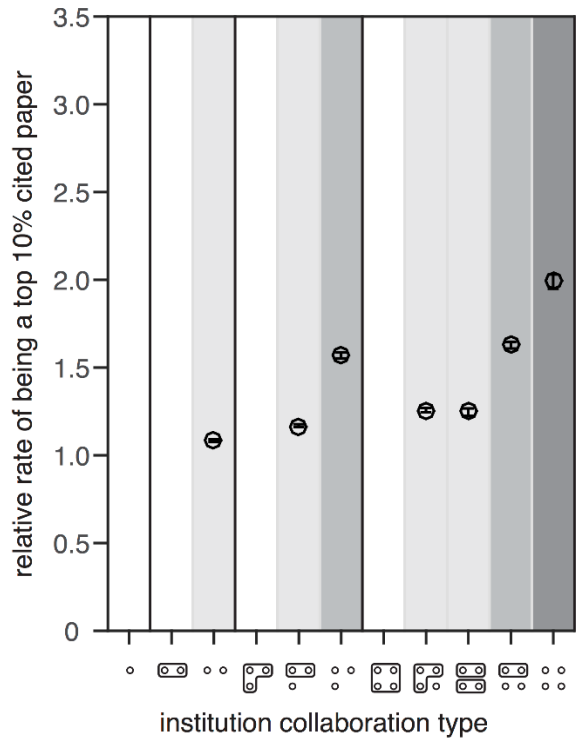

**Fig. S9. Cross-tier institution collaboration correlates with scientific impact.**

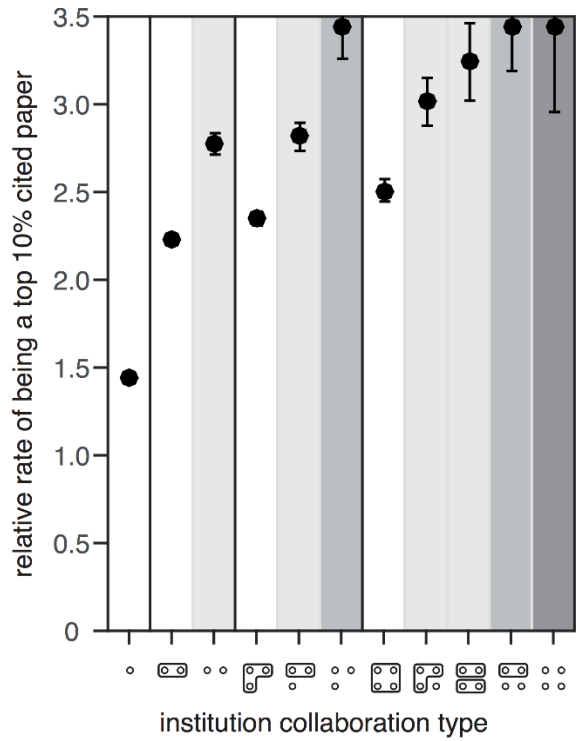

**Fig. S10. Tier-I author collaboration in tier-I institutions correlates with scientific impact.**

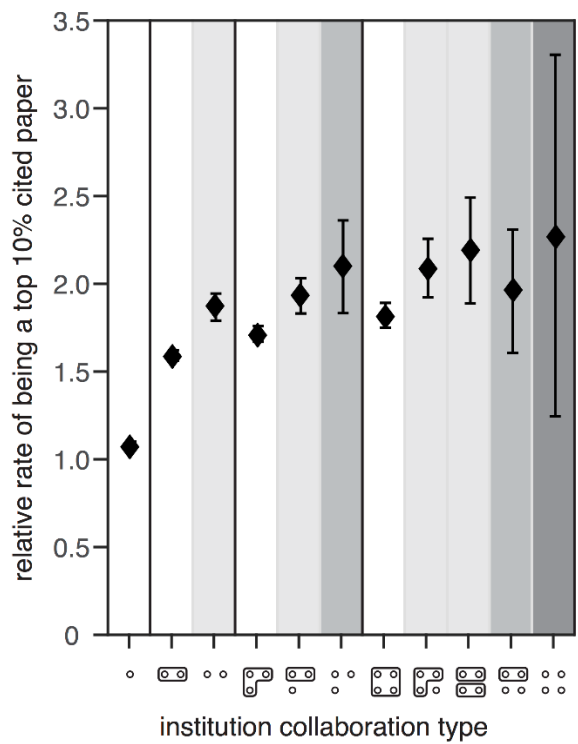

**Fig. S11. Tier-II author collaboration in tier-I institutions correlates with scientific impact.**

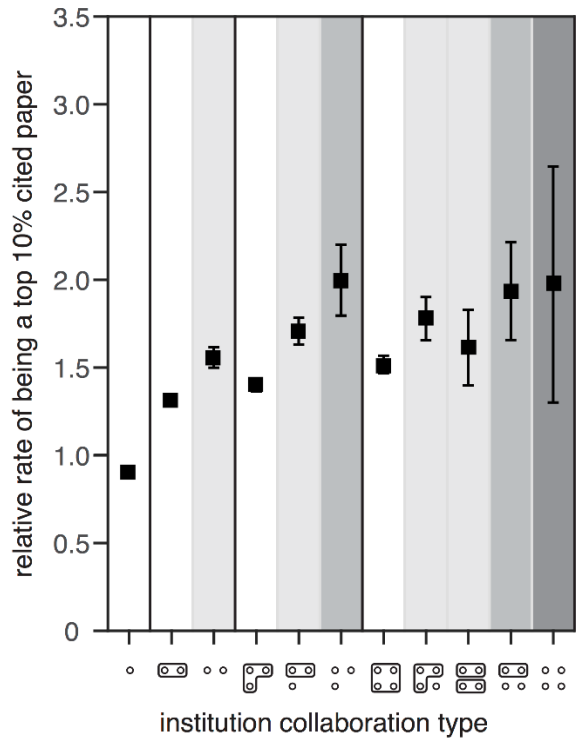

**Fig. S12. Tier-III author collaboration in tier-I institutions correlates with scientific impact.**

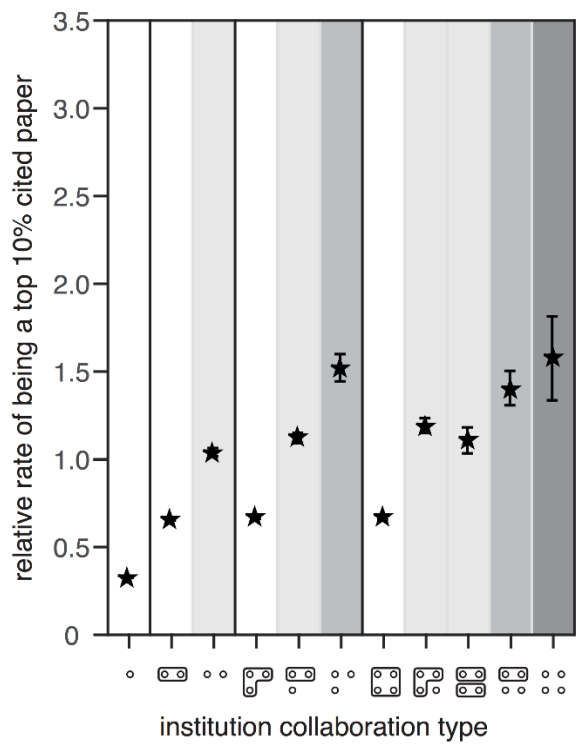

**Fig. S13. Tier-IV author collaboration in tier-I institutions correlates with scientific impact.**

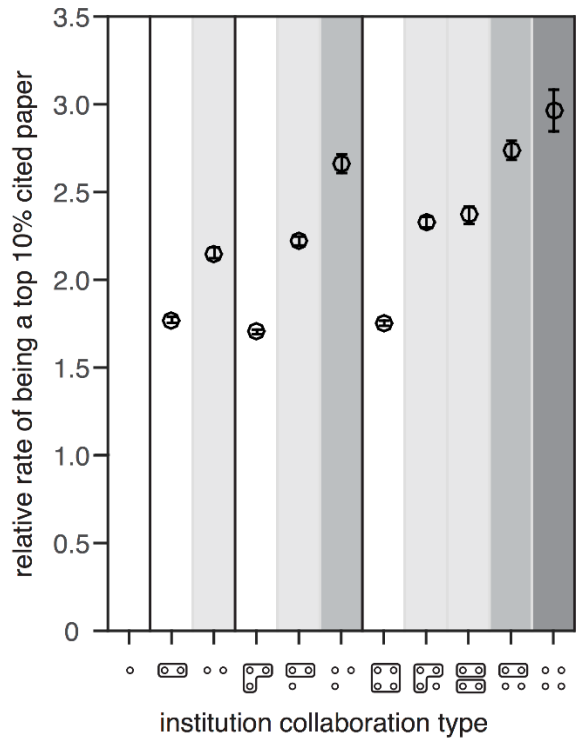

**Fig. S14. Cross-tier author collaboration in tier-I institutions correlates with scientific impact.**

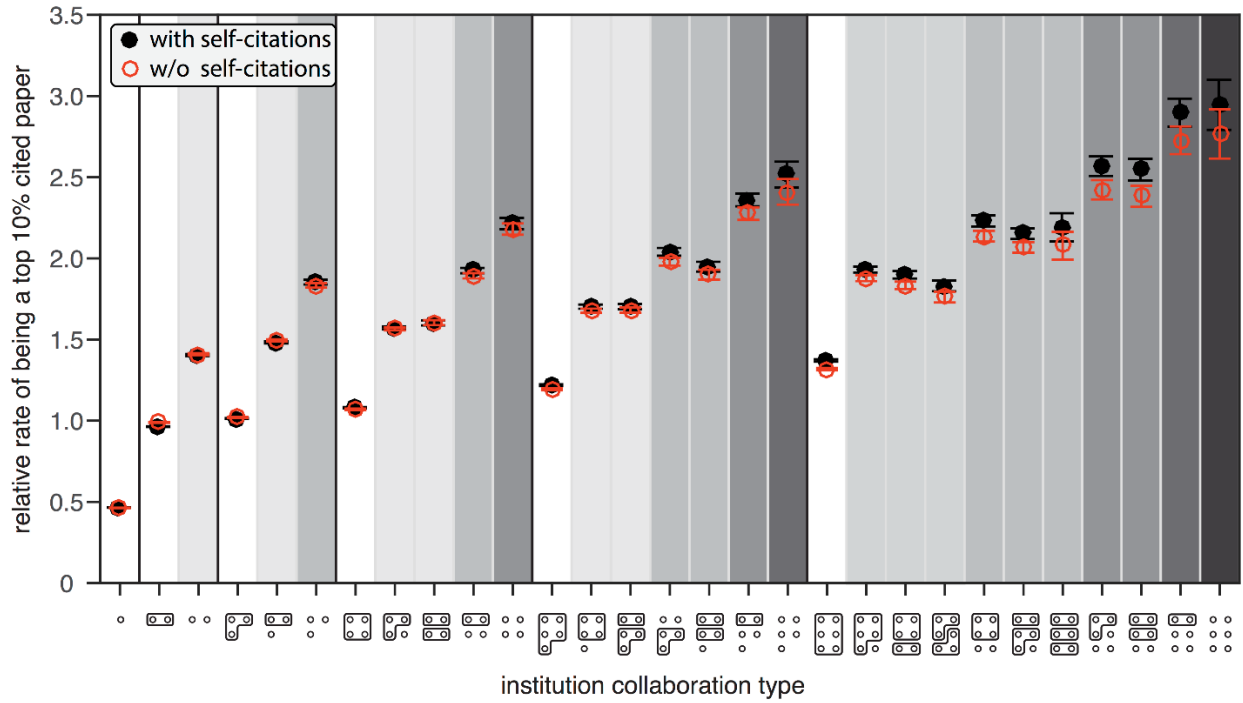

Fig. S15. Institution collaboration diversity correlates with scientific impact with or without self-citations.

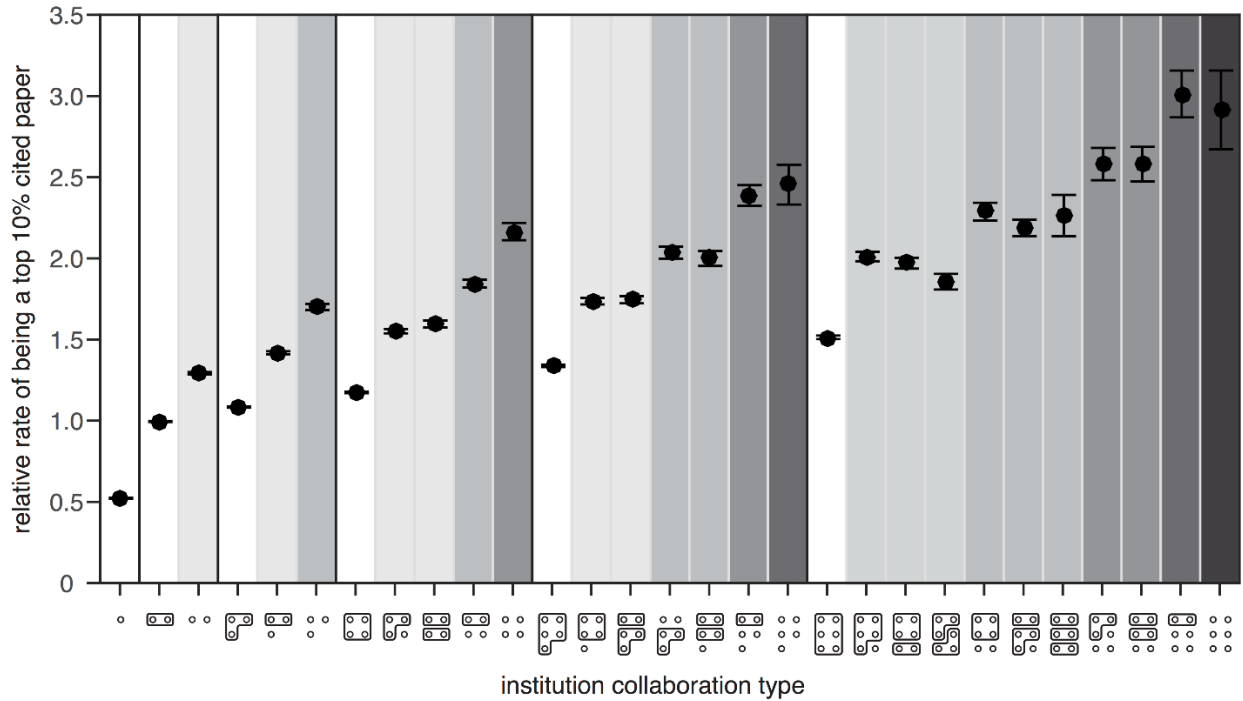

**Fig. S16.** For papers published between year 1965 and 2008, institution collaboration diversity correlates with scientific impact as measured by the citation counts collected within five years.

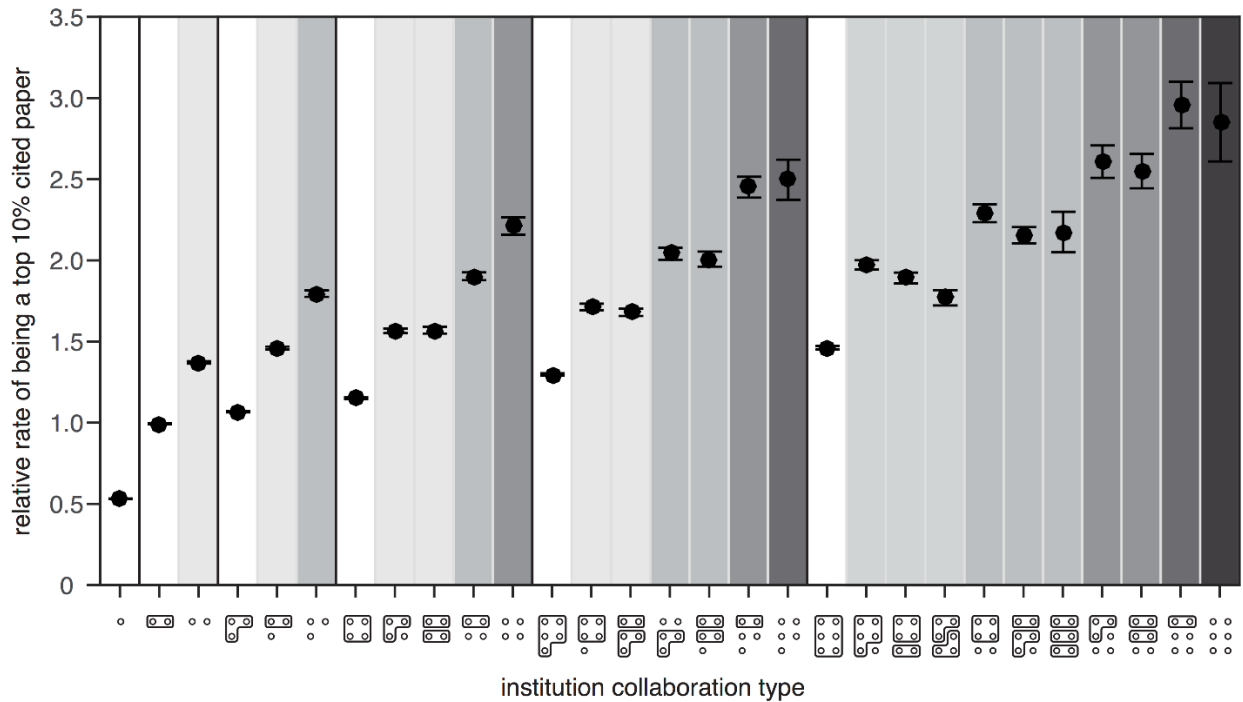

**Fig. S17.** For papers published between year 1965 and 2008, institution collaboration diversity correlates with scientific impact as measured by the citation counts collected within ten years.

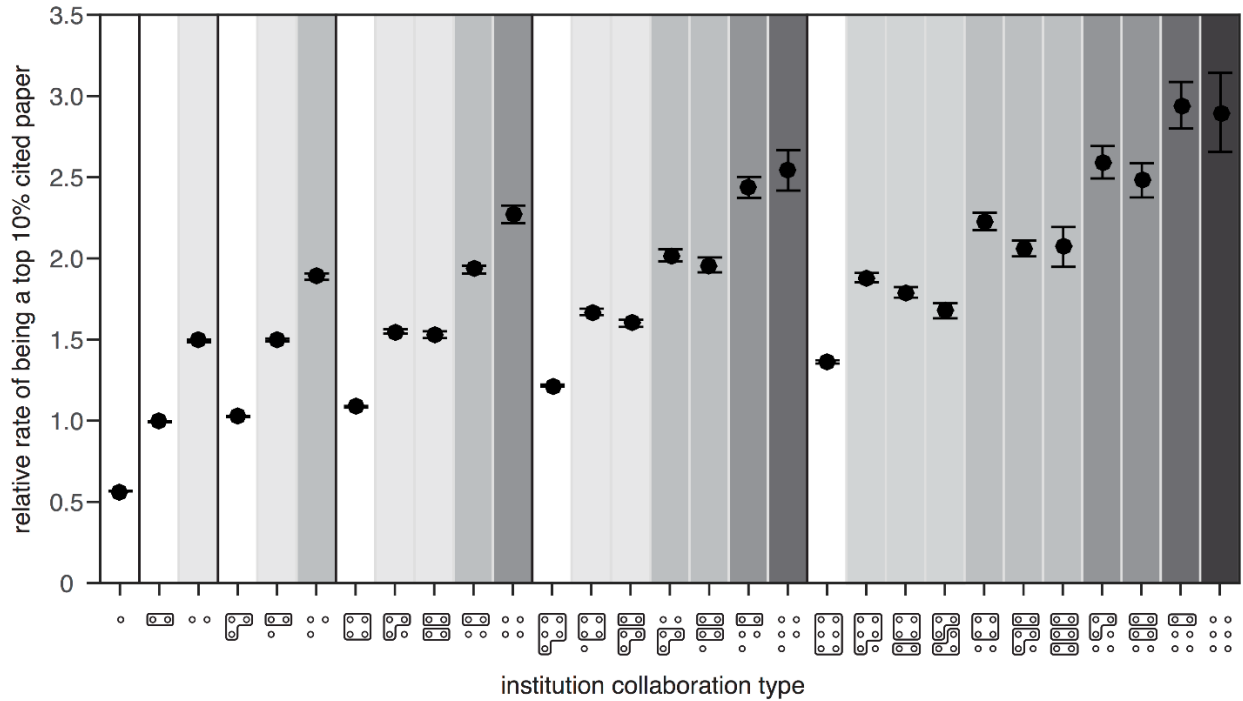

Fig. S18. For papers published between year 1965 and 2008, institution collaboration diversity correlates with scientific impact as measured by the citation counts collected within years to date (2018).

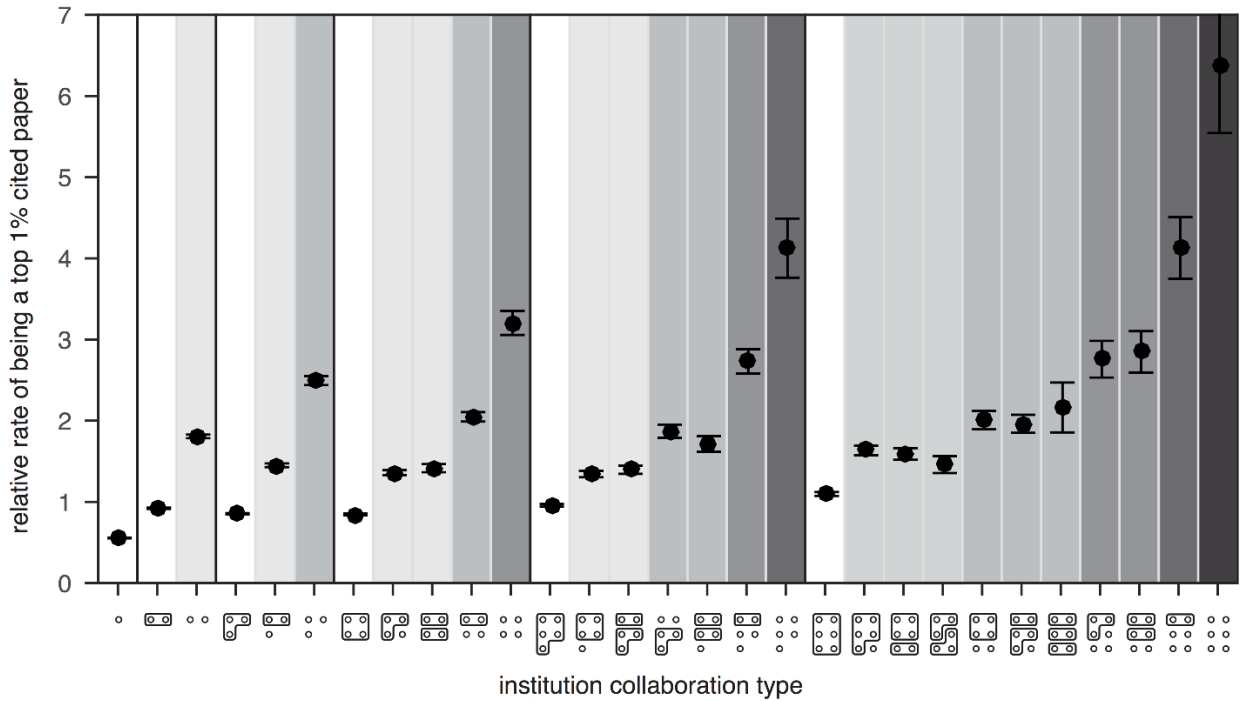

**Fig. S19. Institution collaboration diversity correlates with scientific impact.** The relative rate of being a top 1% cited paper conditioned on institutional collaboration types for one- to six-author teams.

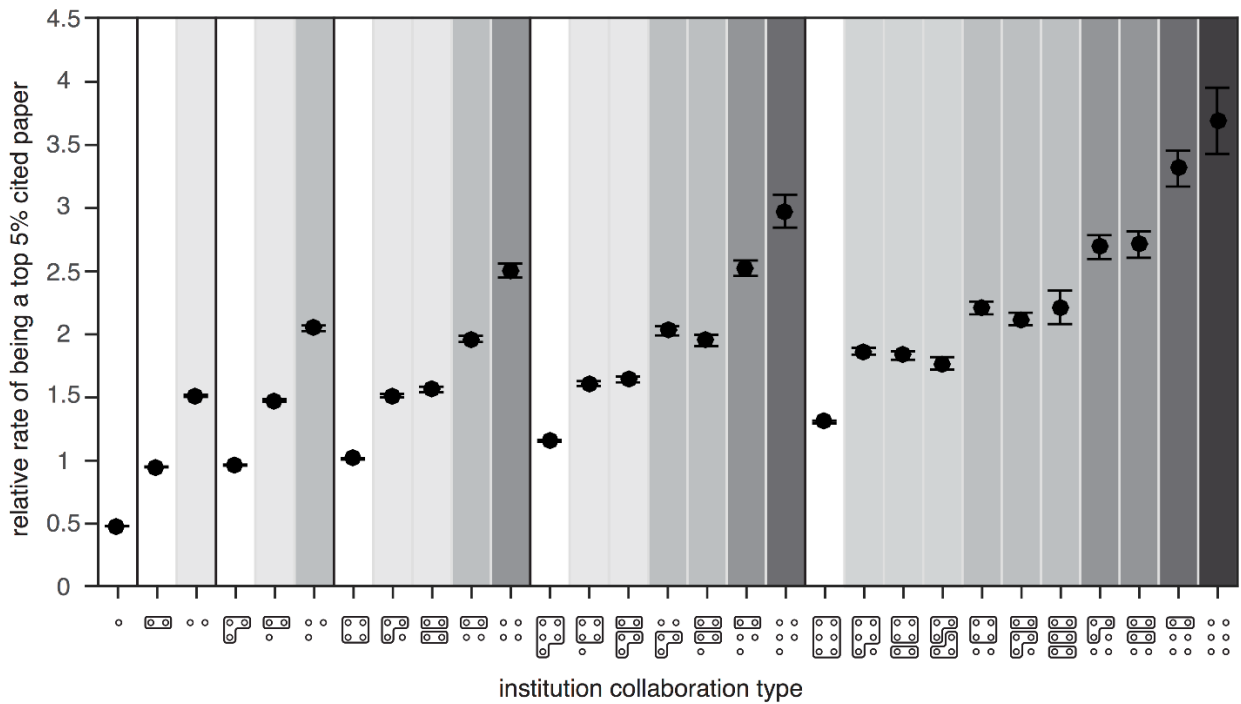

**Fig. S20. Institution collaboration diversity correlates with scientific impact.** The relative rate of being a top 5% cited paper conditioned on institutional collaboration types for one- to six-author teams.

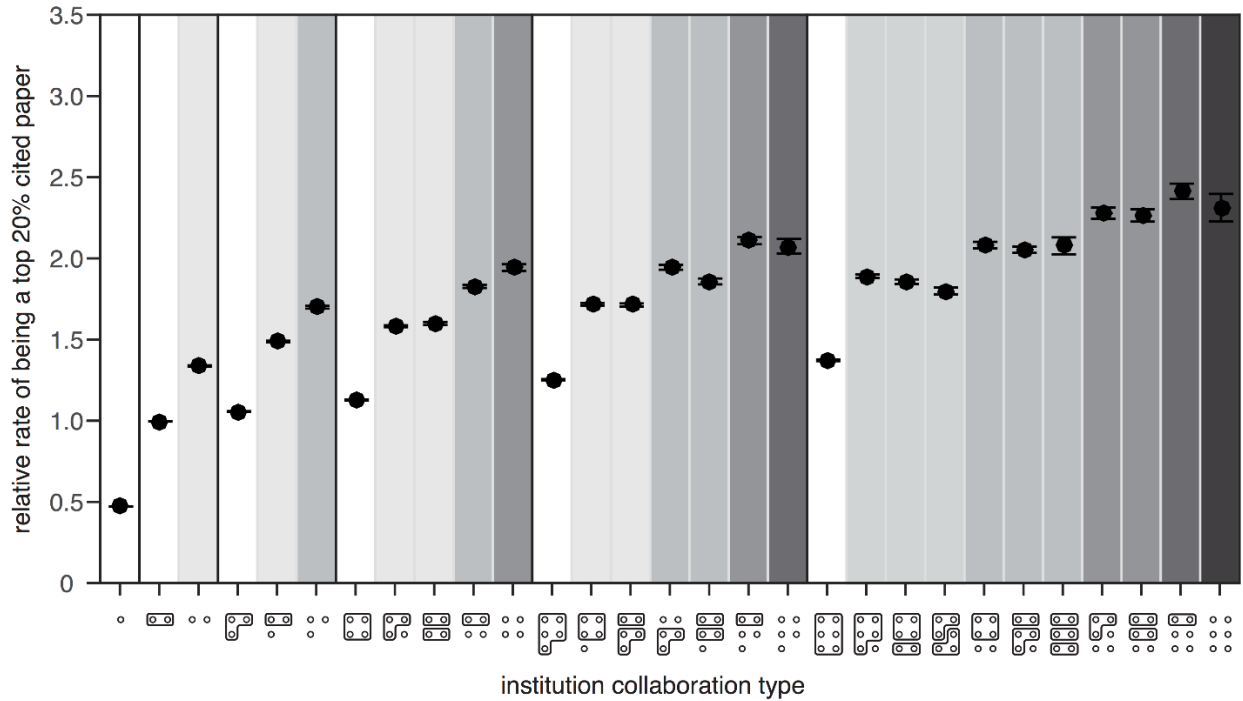

**Fig. S21. Institution collaboration diversity correlates with scientific impact.** The relative rate of being a top 20% cited paper conditioned on institutional collaboration types for one- to six-author teams.

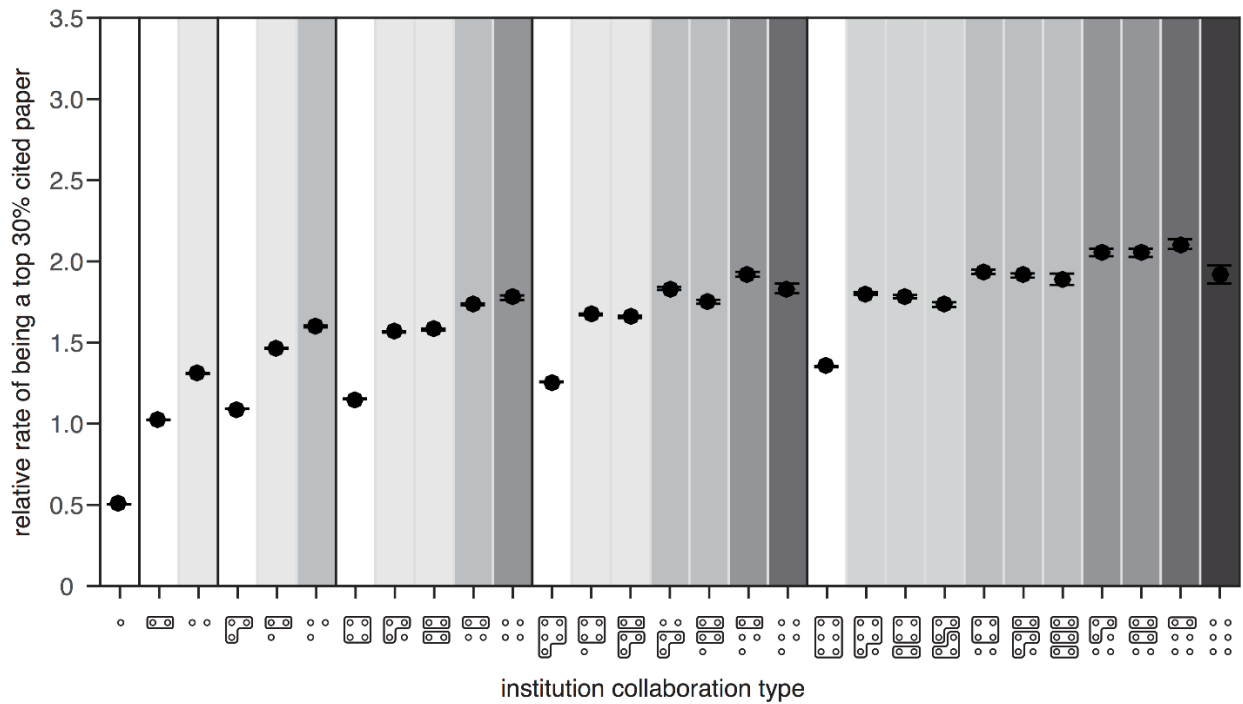

**Fig. S22. Institution collaboration diversity correlates with scientific impact.** The relative rate of being a top 30% cited paper conditioned on institutional collaboration types for one- to six-author teams.

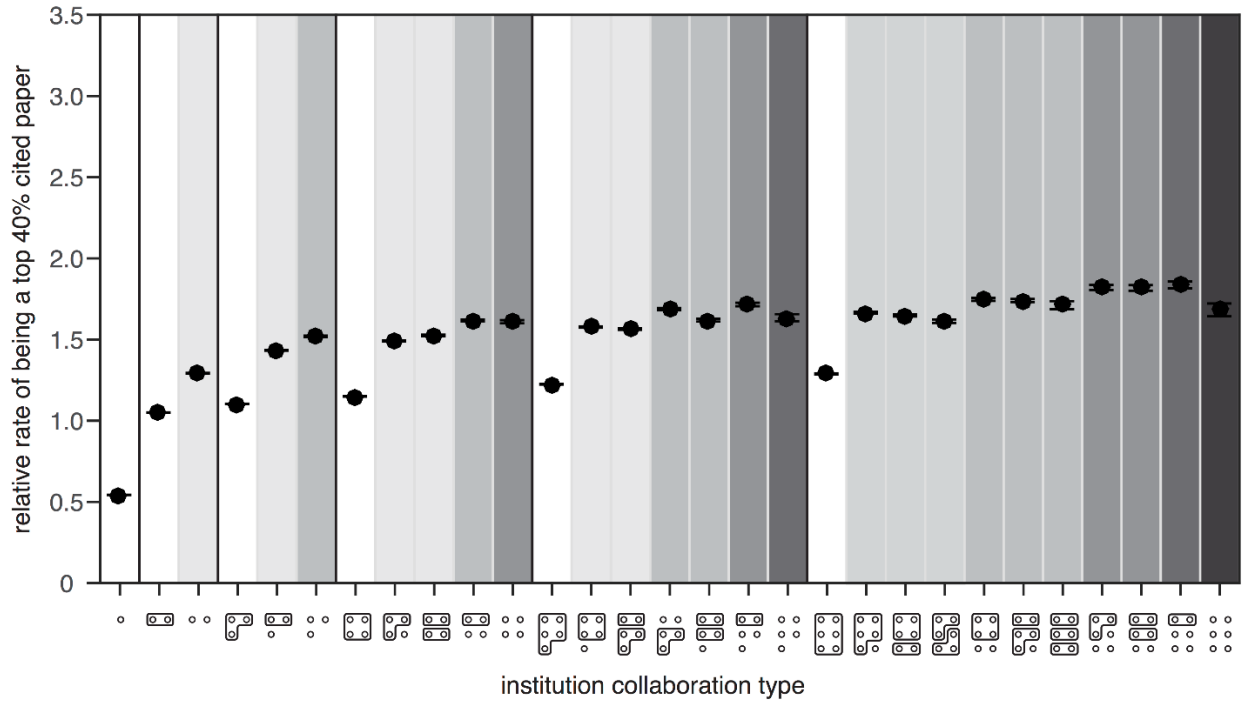

**Fig. S23. Institution collaboration diversity correlates with scientific impact.** The relative rate of being a top 40% cited paper conditioned on institutional collaboration types for one- to six-author teams.

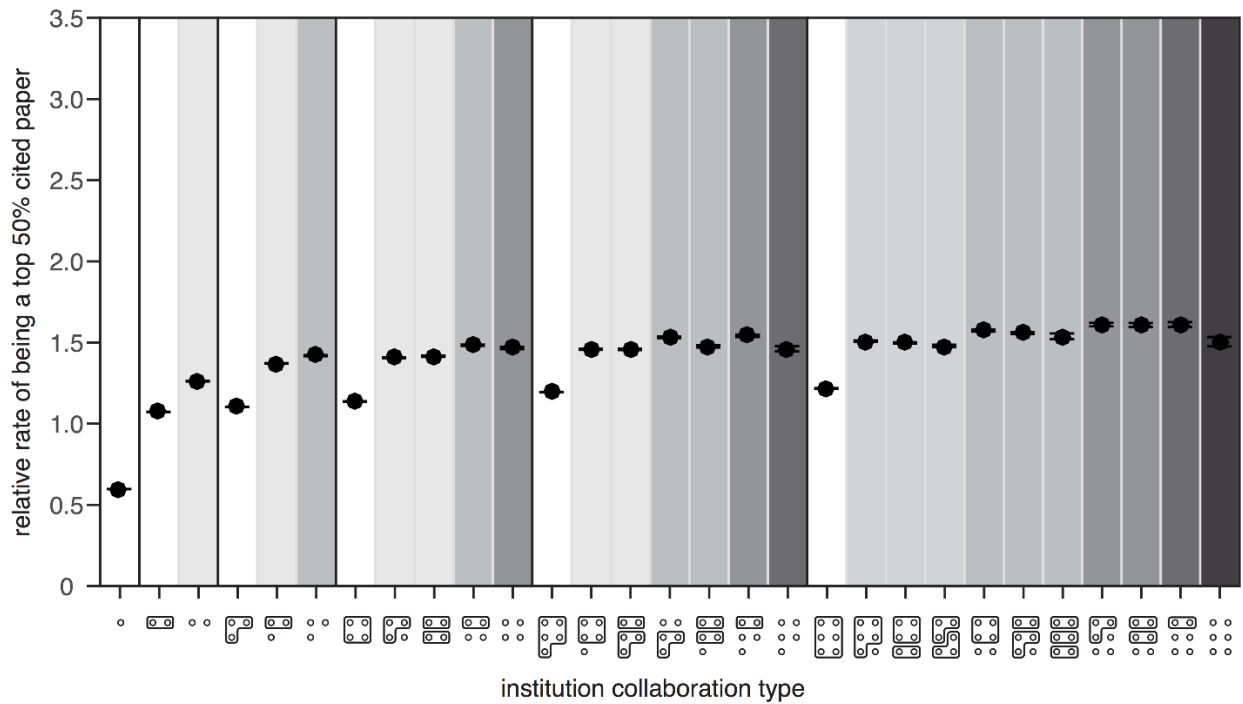

**Fig. S24. Institution collaboration diversity correlates with scientific impact.** The relative rate of being a top 50% cited paper conditioned on institutional collaboration types for one- to six-author teams.